\documentclass{article} 
\usepackage{style/spconf,amsmath,graphicx} 
\usepackage{cite} 
\usepackage{color} 
\usepackage{lipsum} 
\usepackage{amssymb} 
\usepackage{fancyhdr} 
\usepackage[normalem]{ulem} 
\usepackage[version=3]{mhchem} 
\usepackage{tabularx}
\usepackage{graphicx} 
\usepackage{booktabs}
\usepackage{amsmath} 
\usepackage{subfig} 
\usepackage{xcolor} 
\usepackage{colortbl} 
\usepackage{setspace} 
\usepackage{placeins} 
\usepackage{hyperref} 
\usepackage{arydshln}
\usepackage{tikz,pgfplots} 
\usetikzlibrary{calc}
\usepackage{graphicx,pgfplots} 
\usepackage{amsmath} 
\usetikzlibrary{arrows.meta} 
\usepackage{xcolor} 
\usepackage{colortbl} 
\usepackage{placeins}
 \usepgfplotslibrary{groupplots} \usepgfplotslibrary{dateplot} \definecolor{bluegray}{rgb}{0.4, 0.6, 0.8}  
\usetikzlibrary{arrows.meta, shapes.geometric, calc, shadows} 
\usetikzlibrary{arrows, decorations.markings} 
\usetikzlibrary{positioning,arrows,fit} 
\usetikzlibrary{shapes.geometric,backgrounds} 
\usetikzlibrary{shapes.arrows} 
\usepackage{ragged2e}

\title{Automatic dysarthric speech detection exploiting \\ pairwise distance-based convolutional neural networks}

\name{Parvaneh Janbakhshi$^{1,2}$, Ina Kodrasi$^{1}$, Herv\'e Bourlard$^{1,2}$ 
\thanks{The authors would like to acknowledge the support of the Swiss National Science Foundation project no CRSII5\_173711 ``MoSpeeDi'' on ``{\it Motor Speech Disorders: characterizing phonetic speech planning and motor speech programming/execution and their impairments}''.}}

\address{$^{1}$Idiap Research Institute, Speech and Audio Processing Group, Martigny, Switzerland \\
$^{2}$\'Ecole Polytechnique F\'ed\'erale de Lausanne, Lausanne, Switzerland\\
\tt \{parvaneh.janbakhshi,ina.kodrasi,herve.bourlard\}@idiap.ch\\
} 

\begin{document} 
\ninept
\maketitle 
\begin{abstract}
Automatic dysarthric speech detection can provide reliable and cost-effective computer-aided tools to assist the clinical diagnosis and management of dysarthria.
In this paper we propose a novel automatic dysarthric speech detection approach based on analyses of pairwise distance matrices using convolutional neural networks~(CNNs).
We represent utterances through articulatory posteriors and consider pairs of phonetically-balanced representations, with one representation from a healthy speaker (i.e., the reference representation) and the other representation from the test speaker (i.e., test representation).
Given such pairs of reference and test representations, features are first extracted using a feature extraction front-end, a frame-level distance matrix is computed, and the obtained distance matrix is considered as an image by a CNN-based binary classifier.
The feature extraction, distance matrix computation, and CNN-based classifier are jointly optimized in an end-to-end framework.
Experimental results on two databases of healthy and dysarthric speakers for different languages and pathologies show that the proposed approach yields a high dysarthric speech detection performance, outperforming other CNN-based baseline approaches.
 

\end{abstract}


\section{INTRODUCTION} \label{sec:intro}
Dysarthria is a commonly occurring speech disorder arising from brain damage associated with several neurological diseases such as Parkinson's disease (PD) and Amyotrophic Lateral Sclerosis (ALS).
Since several components of the speech production mechanism are disrupted, dysarthria results in impaired phonation, prosody, and articulation~\cite{Darley1969}.
For diagnosis, management, and treatment of patients, these impaired speech dimensions need to be evaluated through clinical auditory-perceptual assessments.
However, such clinical assessments are subjective, time-consuming, and inefficient~\cite{Gavidia1996, baghai2012}.

To assist the clinical diagnosis of dysarthria and to avoid the drawbacks associated with clinical assessments, automatic dysarthric speech detection techniques can be used~\cite{baghai2012}.
Typical automatic techniques developed in the research community are based on i)~handcrafting acoustic features characterizing different impaired speech dimensions and ii) training classifiers using the handcrafted acoustic features to discriminate between dysarthric and healthy speech~\cite{HEGDE2019,GOMEZGARCIA2019}.
Many acoustic features have been exploited to characterize impaired speech dimensions among which are Mel frequency cepstral coefficients, spectro-temporal sparsity parameters and rhythm-based metrics~\cite{Tsanas_ITBE_2012, Orozco-Arroyave_Interspeech_2015, Hemmerling_Interspeech_2016, Sapir_JSLHR_2010, Kodrasi_ICASSP_2019, Kodrasi_ITASLP_2019a, Kodrasi_Interspeech_2020,Hernandez2020}).
Although typical contributions for automatic dysarthric speech detection are based on such features, handcrafted acoustic features may fail to characterize abstract (but similarly important) acoustic cues that can further assist in differentiating dysarthric speech from healthy speech.

Seeking to exploit high-level abstract representations, there has been a growing interest in the research community to leverage data-driven deep learning approaches~\cite{Vasquez2017,Vaiciukynas2018,Kwanghoon,Bhati2019,Mallela2020}.
While deep learning approaches have dramatically improved the state-of-the-art in many speech processing applications, their advantages are yet to be established in the field of pathological speech assessment~\cite{CUMMINS201841}.
The main challenge in successfully exploiting deep learning approaches in pathological speech assessment is alleviating overfitting issues associated with the typically limited training data that is available.

To increase the number of training samples in~\cite{Vasquez2017,Vaiciukynas2018,Kwanghoon}, speech signals are split into short segments (e.g., $160$ ms), each segment is labeled as healthy or dysarthric depending on the label of the complete signal, and convolutional neural networks~(CNNs) are trained on these segments for dysarthric speech detection.
Although considering short segments increases the number of training samples available per speaker, such short segments do not always exhibit dysarthric characteristics, and the CNNs are not guided to ignore speaker variabilities that are unrelated to dysarthria.
A similar approach is also used in~\cite{Mallela2020} where cascaded CNN and long short-term memory~(LSTM) layers are exploited to classify the speech segments.
In~\cite{Bhati2019}, LSTM Siamese networks are used for dysarthric speech detection.
Networks with Siamese architectures are trained on pairs of input data with the same phonetic content.
Pairwise training is advantageous when limited training data is available, since it guides the network to extract features that are discriminative of dysarthria while being robust to other unrelated speaker variabilities.
However, since input data needs to have the same phonetic content, different LSTM networks need to be trained for different utterances.

Instead of using short speech segments to augment training data, we propose to use a CNN-based dysarthric speech detection system exploiting pairwise distance matrices.
While our system benefits from pairwise training, a single network can be used for different utterances since the CNN operates on distance matrices instead of pairs of input data as in~\cite{Bhati2019}.
Inspired by the CNN-based query detection system in~\cite{Ram2020}, we consider utterances from healthy speakers as reference representations, and we propose to compute frame-level distance matrices between these reference representations and phonetically-balanced test representations.
We hypothesize that when the test speaker is healthy, the pattern of the distance matrix between the test and reference (i.e., healthy) representations is different (i.e., it is expected to be more quasi-diagonal) than when the test speaker is dysarthric.
This distance pattern can be used as the input to a CNN-based binary classifier, which then categorizes it as an example from a healthy speaker (i.e., the distance pattern arises from comparing a healthy utterance to the reference representation) or as an example from a dysarthric speaker (i.e., the distance pattern arises from comparing a dysarthric utterance to the reference representation).
Although such a CNN can directly operate on distance matrices computed from user-defined representations of utterances (e.g., the short-time Fourier transform~(STFT)), these user-defined representations might not be optimal for healthy and dysarthric speech detection.
To ensure that distance matrices are computed on optimal representations for our task, we propose to incorporate a front-end feature extraction layer to the network prior to computing distance matrices.
The front-end feature extraction layer, the distance matrix computation, and the final healthy and dysarthric speech detection layers are jointly optimized in an end-to-end learning framework.

For the user-defined utterance representations, we propose to use articulatory posteriors (APs) instead of the STFT representation used in~\cite{Vasquez2017}.
The use of APs is motivated by their potential to characterize articulation deficits in dysarthria, their robustness to noise, and their multilingual and cross-lingual portability~\cite{RASIPURAM2016233}.
Experimental results on two databases of Spanish and French healthy and dysarthric speakers show the advantages of using AP representations in comparison to STFT representations.
Further, experimental results show that the proposed pairwise distance-based CNN with front-end feature extraction can yield a high dysarthric speech detection accuracy, also outperforming a baseline CNN system adapted from~\cite{Vasquez2017} and a pairwise distance-based CNN without front-end feature extraction.


\section{Technical approach}
\label{sec:2}
\begin{figure}[t!]
\resizebox{0.995\linewidth}{!}{
	 \centering {
	 \newcommand{\mygrid}{\tikz{\draw[ draw=black, fill=blue!30,step=1mm,rounded corners=0.2pt] (0,0)  grid (1,1) rectangle (0,0);}}

\tikzset{
  basic box/.style = {
    shape = rectangle,
    align = center,
    draw  = #1,
    fill  = #1!25,
    fill opacity = 0,
    draw opacity = 0,
    rounded corners},
  header node/.style = {
    Minimum Width = header nodes,
    font          = \strut\Large\ttfamily,
    text depth    = +0pt,
    fill          = white,
    draw},
        ncbar angle/.initial=-48,
    ncbar/.style={
        to path=(\tikztostart)
        -- ($(\tikztostart)!#1!\pgfkeysvalueof{/tikz/ncbar angle}:(\tikztotarget)$)
		-- (\tikztotarget)
    },
    ncbar/.default=0.5cm,
  }
 
 \begin{tikzpicture}
    [
       node distance=0.4cm,
       arrow/.style={->,>=stealth,line width=0.4mm,bluegray!70!black, rounded corners=10pt},
       block/.style={rectangle, fill=none, text centered, 
                     rounded corners, draw=bluegray!70!black, line width = 0.6pt, inner sep=-2pt}                      
    ]

    \node[block] (A1)  [inner sep=1pt] {\scriptsize  \begin{tabular}{c}CNN-based classifier\end{tabular}};

    \node (A-s1) [left= 0.3cm of A1]   
{\scriptsize 0/1};    

\node (Grid)  [inner sep=1pt,right= 0.3cm of A1]   
{\mygrid};

    \node (Grid-label)  [above =-0.1cm of Grid] 
{\scriptsize \begin{tabular}{c}Distance matrix\\ (size: S$\times$S)\end{tabular} };    


\node[block] (T)  [draw=green!60!black, inner sep=-0.1pt] [below left= 0.4cm and -0.75cm of Grid]   
{\scriptsize \begin{tabular}{c}Feature\\extraction\end{tabular} };   

    \node (T-label)  [above left=-0.1cm and -1.45cm of T] 
{\scriptsize \begin{tabular}{c}(size: $F_2\times S$)\end{tabular} }; 

\node[block] (R)  [draw=green!60!black, inner sep=-0.1pt] [below right= 0.4cm and 0.3cm of Grid]   
{\scriptsize \begin{tabular}{c}Feature\\extraction\end{tabular} };   

    \node (R-label)  [above right=-0.1cm and -0.9cm of R] 
{\scriptsize \begin{tabular}{c}(size: $F_2\times S$)\end{tabular} }; 

 \node(T-label)  [below=0.2cm of T, inner sep=1pt] {{\color{green!30!black}{\scriptsize \begin{tabular}{c}Test \\ representation\\(size: $F_1\times S$)\end{tabular} }}};       
 
  \node(R-label)  [below=0.2cm of R,inner sep=1pt] {{\color{green!30!black}{\scriptsize \begin{tabular}{c}Reference \\representation\\(size: $F_1\times S$)\end{tabular} }}};       

  \draw[arrow,red!60!black, dashed, dash pattern=on 1pt off 1pt, <->](T.east) to (R.west);
\draw[arrow,red!60!black, ->] (R.north) to [ncbar=0.94cm] (Grid.east);
\draw[arrow,red!60!black, ->] ([xshift=0.5cm]T.north) to  (Grid.south);
\draw[arrow,red!60!black, ->] (Grid.west) to (A1.east);
\draw[arrow,red!60!black, ->] (A1.west) to (A-s1.east);
\draw[arrow,red!60!black, ->] (R-label) to (R.south);
\draw[arrow,red!60!black, ->] (T-label.north) to (T.south);

%
\end{tikzpicture}} }
	\caption{Block diagram of the proposed pairwise distance-based dysarthric speech detection CNN. The two feature extraction blocks share the same set of parameters.}
\label{fig: scheme}
\end{figure}
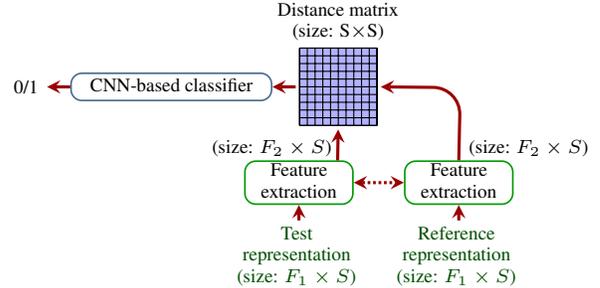

Fig.\ref{fig: scheme} depicts a schematic representation of the proposed pairwise distance-based dysarthric speech detection CNN.
As shown in this figure, the input to the system consists of pairs of reference and test representations of utterances.
We follow the same procedure as in~\cite{Dubagunta2019} to extract AP features for the representations of utterances~(cf. Section~\ref{Reps-setting}).
These representations are transformed through a feature extraction block prior to computing the distance matrix.
The distance matrix is then considered as an image by a CNN-based classifier as in a standard binary image classification task.
The complete architecture is optimized in an end-to-end framework to achieve dysarthric speech detection.

In the following, we present details on the different components of the proposed system, i.e., i) the front-end feature extraction, ii) the distance matrix computation, and iii) the CNN-based classifier.

\subsection{Front-end feature extraction}\label{front-end}

We consider pairs of phonetically-balanced AP representations of utterances from two speakers; one utterance being a reference representation from a healthy speaker and the other utterance being from a test (healthy or dysarthric) speaker.
Let us denote by $\mathbf{R}$ the ($F_1\times M$)--dimensional reference representation, with $F_1$ being the number of AP features and $M$ being the number of time frames in the reference representation.
Similarly, let us denote by $\mathbf{T}$ the ($F_1\times N$)--dimensional test representation, with $N$ being the number of time frames in the test representation.
To be able to handle variable-length inputs, we fix the length of all representations to a predetermined (user-defined) size $S$ as in~\cite{Ram2020}.
Representations with more time frames than $S$, i.e., $M > S$ or $N>S$, are down-sampled by deleting time frames in regular intervals.
Representations with less time frames than $S$, i.e., $M < S$ or $N<S$, are padded at the beginning and end with time frames filled with a constant value.
The constant value is arbitrarily set to the maximum value in the representation.
We denote the resized reference and test representations by $\mathbf{R}_s$ and $\mathbf{T}_s$ and hypothesize that they contain similar (healthy or dysarthria-related) cues as in the original representations $\mathbf{R}$ and $\mathbf{T}$.

The front-end feature extraction block transforms the ($F_1 \times S$)--dimensional representations $\mathbf{R}_s$ and $\mathbf{T}_s$ into ($F_2 \times S$)--dimensional representations.
To this end, we use a 1D convolution layer with $F_2$ channels such that the $F_1$--dimensional AP feature vectors for each time frame are transformed into $F_2$--dimensional feature vectors.
Since this layer is jointly optimized with the distance matrix computation~(cf. Section~\ref{DistMat}) and the CNN-based classifier~(cf. Section~\ref{CNN1}) in an end-to-end framework, it can be expected that the transformed ($F_2 \times S$)--dimensional representations are more discriminative representations for the dysarthric speech detection task.

The architecture of the front-end layer is summarized in Table~\ref{table0-1}, where we have used $F_2 = 32$.
It should be noted that the parameters of the front-end feature extraction layer to compute both test and reference feature representations are the same (cf. Fig.~\ref{fig: scheme}).

\begin{table}[b!] \centering 
	 \caption{{\it Front-end feature extraction architecture.\vspace{-0.2cm}}} \label{table0-1}
	{\fontsize{7}{7}\selectfont
	\begin{tabularx}{\linewidth}{X|c} \\\toprule
		Layer & Description      \\\midrule\midrule
		Input& Size: ($1$x$F_1$x$S$): input speech representation\\
		Conv1d + Relu & Channel: in=$1$, out=$32$, Filter: $F_1$x$1$, Stride: $1$\\\bottomrule
	\end{tabularx}}
\end{table}

\subsection{Distance matrix computation}
\label{DistMat}

The distance matrix is computed from the representations at the output of the feature extraction block.
Let us denote the reference representation after feature extraction by $\hat{\mathbf{R}} = [\mathbf{r}_1, \ldots, \mathbf{r}_S]$, with $\mathbf{r}_i$, $i = 1, \ldots, S$, being the $F_2$--dimensional feature vector at time frame~$i$.
Similarly, the test representation after feature extraction is denoted by $\hat{\mathbf{T}} = [\mathbf{t}_1, \ldots, \mathbf{t}_S]$, with $\mathbf{t}_j$, $j = 1, \ldots, S$, being the $F_2$-dimensional feature vector at time frame~$j$.
The frame-level distance matrix $\mathbf{D}$ between the representations $\hat{\mathbf{T}}$ and $\hat{\mathbf{R}}$ is an $(S \times S)$--dimensional matrix, where the $(i,j)$--th entry is computed as the distance $d$ between $\mathbf{t}_i$ and $\mathbf{r}_j$, i.e.,
\begin{equation}
  \mathbf{D}_{i,j} =d(\mathbf{t}_i, \mathbf{r}_j). \label{eq1}
\end{equation}
To compute $\mathbf{D}$ within the proposed end-to-end framework, Euclidean distance is used, i.e., $d(\mathbf{t}_i,\mathbf{r}_j)=||\mathbf{t}_i-\mathbf{r}_j||$.
Since the reference representation $\hat{\mathbf{R}}$ always belongs to a healthy speaker, we expect the pattern of the so-computed distance matrix $\mathbf{D}$ to be more quasi-diagonal (i.e., contain more zeros on the diagonal due to similar $\mathbf{t}_i$ and $\mathbf{r}_j$) when the test representation $\hat{\mathbf{T}}$ belongs to a healthy speaker than when it belongs to a dysarthric speaker.

\subsection{CNN-based classifier with pairwise distance matrices}\label{CNN1}
The distance matrices computed in Section~\ref{DistMat} serve as input to our CNN classifier.
As summarized in Table~\ref{table0}, the CNN classifier consists of two 2D convolutional layers, followed by two Maxpooling and two fully connected (FC) layers.
To prevent overfitting, dropout is employed during training.
The label for each distance matrix fed into the CNN classifier is the label of the test speaker (healthy or dysarthric) used for the distance matrix computation.

\begin{table}[b!] \centering 
	 \caption{{\it Architecture of the proposed CNN-based classifier operating on pairwise distance matrices.\vspace{-0.2cm}}} \label{table0}
	{\fontsize{7}{7}\selectfont
	\begin{tabularx}{\linewidth}{X|c} \\\toprule
		Layer & Description      \\\midrule\midrule
		Input& Size: ($1$xSx$S$) input distance matrix \\
		Conv2d + Relu & Channel: in=$1$, out=$16$, Filter: $10$x$10$, Stride: $1$ \\
		Maxpool2d & Channel: in=$16$, out=$16$, Filter: $2$x$2$, Stride: $2$ \\
		Conv2d + Relu & Channel: in=$16$, out=$16$, Filter: $10$x$10$, Stride: $1$\\
		Maxpool2d & Channel: in=$16$, out=$16$, Filter: $2$x$2$, Stride: $2$ \\
		Dropout & Probability: $0.5$ \\
		FC + Relu & Input: $784$, Output: $128$ \\
		FC + Softmax & Input: $128$, Output: $2$  \\\bottomrule
	\end{tabularx}}
\end{table}

The classifier is trained using distance matrices computed from all phonetically-matched pairs of test and reference representations in the training set.
As mentioned in Section~\ref{sec:intro}, a single network can be used for different utterances since the CNN operates on distance matrices instead of pairs of input data as in~\cite{Bhati2019}.
To evaluate an utterance from an unseen test speaker, we pair it to its phonetically-matched counterpart from many reference speakers in the training set and compute multiple distance matrices.
All available distance matrices are then independently processed by the CNN classifier, and the final decision for the unseen test speaker is made by applying soft voting on all CNN prediction scores for all available distance matrices from that speaker.

\section{Experimental results}
\label{sec: exp}
In this section, the performance of the proposed pairwise distance-based dysarthric speech detection system is evaluated and compared to baseline systems.
\subsection{Databases} \label{subsec:data}

To investigate the applicability and generalisability of the proposed approach to different pathologies and languages, two databases are considered.

{\it PC-GITA database~\cite{GITA}.}\enspace We consider Spanish recordings from $50$ PD patients ($25$ males, $25$ females) and $50$ healthy speakers ($25$ males, $25$ females) from the PC-GITA database~\cite{GITA}.
Each speaker utters $24$ words, which are recorded at a sampling frequency of $44.1$~kHz.
After downsampling to $16$~kHz, speech-only segments are manually extracted from the recordings.

{\it MoSpeeDi database.}\enspace We consider French recordings from $20$ PD and ALS patients ($14$ males, $6$ females) and $20$ healthy speakers ($10$ males, $10$ females) from Geneva University Hospitals and University of Geneva.
Each speaker utters $54$ pseudo-words, which are recorded at a sampling frequency of $44.1$~kHz.
After downsampling to $16$~kHz, speech-only segments are extracted from the recordings using an energy-based voice activity detector~\cite{praat}.

\subsection{Articulatory posterior representation}\label{Reps-setting}
AP representations are extracted as in~\cite{Dubagunta2019}, where frame-level posteriors of four articulatory categories are computed, i.e., manner of articulation (e.g., degree of constriction), place of constriction, height of the tongue, and vowel.
Posteriors for each category are estimated using CNNs trained on healthy speech data from the AMI corpus~\cite{Carletta2005} based on acoustic phoneme-to-articulatory feature mappings~\cite{RASIPURAM2016233}.
By concatenating all extracted APs, $F_1=53$ features per time frame are obtained.
For details on the training procedure for AP feature extraction, the reader is referred to~\cite{Dubagunta2019}.

\subsection{Baseline networks}\label{baseline}

To demonstrate the advantages of the proposed approach, the following two baseline systems B-CNN$_1$ and \mbox{B-CNN$_2$} are considered.

\emph{B-CNN$_1$.} \enspace We have implemented a baseline CNN adapted from~\cite{Vasquez2017}, which is trained on log magnitude of STFT representations of short (i.e., $160$~ms) segments of speech with 50\% overlap.
The STFT representations are computed using $10$ ms Hanning windows without overlap, resulting in $129$ frequency bins for each time frame.
The final decision for an unseen speaker is made by applying soft voting on the segment-level CNN prediction scores.
To demonstrate the advantage of using AP representations instead of STFT, such a baseline CNN is also trained on the logarithm of AP representations.
The architecture of this baseline system is summarized in Table~\ref{table1}.

\emph{B-CNN$_2$.} \enspace To further establish the advantages of the proposed end-to-end CNN framework (which uses a front-end feature extraction layer), a second baseline is implemented where the proposed CNN-based classifier in Section~\ref{CNN1} is trained on distance matrices computed directly from AP representations (i.e., without using the front-end feature extraction layer).
To compute such distance matrices, Kullback-Leibler divergence is used as the local distance measure in~\eqref{eq1}.
The architecture of this baseline system is the same as in Table~\ref{table0}.


\begin{table}[b!] \centering 
	 \caption{{\it Architecture of the baseline B-CNN$_1$ adapted from~\cite{Vasquez2017}.\vspace{-0.2cm}}} \label{table1}
	{\fontsize{7}{7}\selectfont
	\begin{tabularx}{\linewidth}{X|c} \\\toprule
		Layer & Description      \\\midrule\midrule
		Input & Size: ($1$xFx$16$); F: dimension of input representation \\
		Conv1d + Relu & Channel: in=$1$, out=$32$, Filter: $F$x$1$, Stride: $1$ \\
		Conv1d + Relu & Channel: in=$32$, out=$16$, Filter: $1$x$4$, Stride: $1$\\
		Dropout & Probability: $0.5$ \\
		FC + Relu & Input: $208$, Output: $128$ \\
		FC + Softmax & Input: $128$, Output: $2$  \\\bottomrule
	\end{tabularx}}
\end{table}

\subsection{Training and evaluation}\label{training}

The validation strategy on the PC-GITA and MoSpeeDi databases is a stratified speaker-independent 10-fold and 5-fold cross-validation framework, respectively (i.e., speakers in each fold are different).
In each training fold, a development fold with the same size as the test fold is set aside for early-stopping.
\mbox{Z-score} normalization is applied to all input representations.
All networks are trained using the stochastic gradient descent (SGD) algorithm and the cross-entropy loss.
The batch size is $256$, and the initial learning rate is $0.05$.
The learning rate is divided by $5$ each time the loss on the development set does not decrease for $5$ consecutive iterations.
The training is stopped either after $100$ epochs or after the learning rate has reached the value $10^{-6}$.
 
Random weight initialization is used for the baselines B-CNN$_1$ and B-CNN$_2$.
The weights on the first convolution layer of the trained baseline B-CNN$_1$ are used to initialize the front-end feature extraction layer of the proposed end-to-end CNN.
The weights of the trained baseline \mbox{B-CNN$_2$} are used to initialize the classifier layers of the proposed end-to-end CNN.

The number of total samples (training/testing) available for the different considered networks is as follows.
Using the STFT representation for B-CNN$_1$ results in $17383$ (PC-GITA) and $25197$ (MoSpeeDi) segments.
Using the AP representation for B-CNN$_1$ results in $17368$ (PC-GITA) and $25907$ (MoSpeeDi) segments.
The number of distance matrices computed from all pairs of reference and test AP representations for B-CNN$_2$ and the proposed CNN is $96000$ (PC-GITA) and $25920$ (MoSpeeDi).

The dysarthric speech detection performance is evaluated in terms of the area under ROC curve (AUC).
In addition, we also compute the classification accuracy using a decision threshold of $0.5$.
To reduce the impact of the random seed on the final model parameters, we have trained all networks with $3$ different random seeds.
The reported performance measures are the mean and standard deviation of the performance obtained by models trained using different seeds.

\begin{table}
	[b!] \centering 
	\caption{{\it Mean and standard deviation of the AUC score and classification accuracy [\%] using the baseline B-CNN$_1$ with STFT and AP representations on the PC-GITA and MoSpeeDi databases.\vspace{-0.2cm}}} \label{table2}
	{\fontsize{7.2}{8}\selectfont
	\begin{tabularx}{\linewidth}{XX|rr}\\ \toprule
		Database & Input representation& AUC & Accuracy\\\midrule\midrule
		Spanish PC-GITA & STFT & $0.56\pm0.03$ & $53.67\pm3.29$ \\
		Spanish PC-GITA & AP & $0.75\pm0.00$ & $72.00\pm0.81$ \\\midrule
		French MoSpeeDi & STFT & $0.64\pm0.02$ & $52.50\pm0.00$\\
		French MoSpeeDi & AP & $0.73\pm0.03$ & $60.83\pm3.12$ \\\bottomrule
	\end{tabularx}}
\end{table}
   
\subsection{Results}

Table~\ref{table2} presents the AUC and classification accuracy values obtained using B-CNN$_1$ on STFT and AP representations for both considered databases.
It can be observed that the AP representation yields a better performance than the STFT on both databases, with a particularly significant improvement observed for the PC-GITA database.
These results are to be expected given the advantages of articulatory modeling of speech using AP as described in Section~\ref{sec:intro}. 

It should be noted that the CNN proposed in~\cite{Vasquez2017} was trained on the PC-GITA database using speech segments centered at transitions between voiced and unvoiced regions.
However, although not presented here due to space constraints, using such segments did not result in a better performance than the performance presented in Table~\ref{table2}.
Further, it should be noted that ~\cite{Vasquez2017} uses more recordings than the word recordings we have used here.
To ensure that the conclusions derived in this paper on the advantages of the proposed approach as opposed to B-CNN$_1$ are still valid even when more recordings are available for use in B-CNN$_1$, we have investigated the performance of B-CNN$_1$ using AP representations on both databases when all available recordings are used (rather than just words).

Using all available recordings and the AP representation for \mbox{B-CNN$_1$} results in $74762$ (PC-GITA) and $54626$ total available segments.
In this case, B-CNN$_1$ yields AUC and accuracy values of $0.78$ and $73.33\%$ on the PC-GITA database and $0.75$ and $60.00\%$ on the MoSpeeDi database.
When comparing these results to the ones obtained using only word recordings (cf. entries for AP representations in Table~\ref{table2}), we observe that increasing the used speech material does not significantly improve the dysarthric speech detection performance of B-CNN$_1$.
%
In summary, the presented results demonstrate the advantage of using AP representations as opposed to the STFT representations used in~\cite{Vasquez2017}.
In the following, the performance of both baseline systems B-CNN$_1$ and B-CNN$_2$ and of the proposed end-to-end CNN is compared when AP representations are used.

Table~\ref{table3} presents the AUC and classification accuracy values of the baseline systems B-CNN$_1$ and B-CNN$_2$ and of the proposed approach on both databases.
Bold entries indicate the maximum performance for each database.
It can be observed that the proposed pairwise distance-based CNN with front-end feature extraction outperforms both considered baselines in terms of both performance measures on both databases.
Comparing the difference in performance between the proposed framework and B-CNN$_2$ shows that incorporating a feature extraction front-end significantly improves the performance in comparison to computing distance matrices directly on AP representations.
Analyzing the learned representations from the feature extraction front-end remains a topic for future investigation.


In summary, the presented results show that the proposed pairwise distance-based CNN with a front-end feature extraction layer is successfully applicable to the dysarthric speech detection task.
Although a small number of utterances per speaker are used, the proposed approach outperforms baseline systems for different databases with different languages and disorders.

\begin{table}
	[t!] \centering 
	\caption{{\it Mean and standard deviation of the AUC score and classification accuracy [\%] using the baseline B-CNN$_1$ and B-CNN$_2$ and the proposed pairwise distance-based approach with a front-end feature extraction layer on the PC-GITA and MoSpeeDi databases.}\vspace{-0.2cm}} \label{table3}
	{\fontsize{7.5}{8}\selectfont
	\begin{tabularx}{\linewidth}{XX|ll}\\ \toprule
		Database & CNN & AUC & Accuracy \\\midrule\midrule
		Spanish PC-GITA & Baseline B-CNN$_1$ & $0.75\pm0.00$ & $72.00\pm0.81$ \\
		Spanish PC-GITA & Baseline B-CNN$_2$ & $0.78\pm0.01$ & $68.33\pm0.74$ \\
		Spanish PC-GITA & Proposed & $\mathbf{0.83}\pm0.01$ & $\mathbf{77.67}\pm0.47$ \\\midrule
		French MoSpeeDi & Baseline B-CNN$_1$ & $0.73\pm0.03$ & $60.83\pm3.11$  \\
		French MoSpeeDi & Baseline B-CNN$_2$ & $0.77\pm0.00$ & $70.83\pm2.35$ \\
		French MoSpeeDi & Proposed & $\mathbf{0.84}\pm0.02$ & $\mathbf{76.67}\pm4.25$ \\\bottomrule
	\end{tabularx}}
\end{table}

\section{Conclusion} \label{sec:conclusion} 

In this study, we explored the feasibility of automatic dysarthric speech detection using a pairwise distance-based CNN.
The proposed approach compares frame-level distance patterns between phonetically-balanced AP representations from healthy (i.e., reference) and test speakers.
After extracting features from such representations and processing their distance matrix, a CNN-based classifier predicts whether the test representation is from a healthy or dysarthric speaker.
Feature extraction, distance matrix computation, and classification are jointly optimized in an end-to-end framework.
Experimental results on two dysarthric speech databases have shown that the proposed approach is generalizable across languages, obtaining a high dysarthric speech detection accuracy and outperforming state-of-the-art CNN-based systems.


\bibliographystyle{style/IEEEtran}
{
\footnotesize 
 \bibliography{ref/refs.bib, refs}
}

\end{document}